\documentclass{aa}

\usepackage{graphicx}

\def\jcd{Christensen-Dalsgaard}
\newlength{\figwidth}
\newcommand{\vort}{\mbox{\boldmath $\omega$}}
\newcommand{\curl}{\mbox{\boldmath $\nabla \times$}}
\newcommand{\uvr}{\mbox{\boldmath $\hat{\bf e}_r$}}
\def\p{\partial}
\def\rb{\bar{\rho}}
\def\rbi{\frac{1}{\bar{\rho}}}
\def\nab{\mbox{\boldmath $\nabla$}}
\def\Om{\mbox{\boldmath $\Omega_0$}}

\begin{document}

\title{Is the solar convection zone in strict thermal wind balance?}
\titlerunning{Is the solar convection zone in strict thermal wind balance}

\author{A. S. Brun \inst{1,2}, H. M.~Antia \inst{3}, \and S. M.~Chitre \inst{4}}
\authorrunning{Brun, Antia \& Chitre}
\offprints{A. S. Brun \\ \email{sacha.brun@cea.fr}}

\institute{Laboratoire AIM, CEA/DSM-CNRS-Universit\'e Paris Diderot,
IRFU/SAp, 91191 Gif sur Yvette, France \and
LUTH, Observatoire de Paris, CNRS-Universit\'e Paris Diderot, Place Jules Janssen, 92195 Meudon, France \and 
Tata Institute of Fundamental Research, Homi Bhabha Road,
   Mumbai 400005, India \and
Centre for Basic Sciences, University of Mumbai, Mumbai 400098, India}

\date{Received ?? ; accepted ??}

\abstract{The solar rotation profile is conical rather than cylindrical as one could expect
from classical rotating fluid dynamics (e.g. Taylor-Proudman theorem). Thermal coupling to the tachocline, baroclinic effects
and latitudinal transport of heat have been advocated to explain this peculiar state of rotation.}
{To test the validity of thermal wind balance in the solar
convection zone using helioseismic inversions for both the angular velocity and fluctuations in entropy
and temperature.}
{Entropy and temperature fluctuations obtained from 3-D hydrodynamical numerical simulations
of the solar convection zone are compared with solar profiles obtained from helioseismic inversions.}
{The temperature and entropy fluctuations in 3-D numerical simulations have smaller amplitude in the bulk of the solar convection zone than those found from seismic inversions. 
Seismic inversion find variations of temperature from about 1 K  at the surface up to 100 K at the base of the convection zone while in 3-D simulations 
they are of order 10 K throughout the convection zone up to 0.96 $R_{\odot}$. In 3-D simulations, baroclinic effects are found to be important to tilt the isocontours of $\Omega$ 
away from a cylindrical profile  in most of the convection zone helped by Reynolds and viscous stresses at some locations. 
By contrast the baroclinic effect inverted by helioseismology are much larger than what is required to yield the observed angular velocity profile.}
{The solar convection does not appear to be in strict thermal wind balance, Reynolds stresses 
must play a dominant role in setting not only the equatorial acceleration but also the observed conical angular velocity profile.}


\maketitle

\section{Introduction}

Helioseismic data from the Global Oscillation Network Group (GONG)
and the Michelson Doppler Imager (MDI) have been used to infer the
rotation profile in the solar interior (e.g., Thompson et al.~1996; Schou et al.~1998).
The inversion results show that isocontours of the differential rotation $\Omega(r,\theta)$ are conical
at mid-latitude, rather than cylindrical as was expected from early
numerical simulations (e.g., Glatzmaier \& Gilman 1982; Gilman \& Miller 1986).
More recent theoretical work (Durney 1999; Kitchatinov \& Rudiger 1995;
Brun \& Toomre 2002 (hereafter BT02); Rempel 2005; Miesch et al.~2006 (hereafter MBT06); Brun \& Rempel 2008; Balbus et al. 2009)
indicate that in order to break the Taylor-Proudman constraint of
cylindrical $\Omega$, the Sun must either have a systematic latitudinal heat transfer in
its convection zone or thermal forcing from the tachocline or most likely both.
This is due to the so-called thermal wind balance (Pedlosky 1987), i.e., the existence in the solar convection zone of 
latitudinal entropy (or temperature) variation due to baroclinic
effect  can result in a rotation state that breaks the Taylor-Proudman constraint.
Such latitudinal variations of the thermal properties at the solar surface have been looked for
observationally by several groups since the late 60's (e.g., Dicke \& Goldenberg 1967; 
Altroch \& Canfield 1972; Koutchmy et al.~1977; Kuhn et al.~1985, 1998, Rast et al.~2008; 
to cite only a few). This is a difficult task since one has to correct for limb
darkening effect, photospheric magnetic activity, instrument bias and many other subtle effects 
to extract a relatively weak signal (see Rast et al.~2008). In most cases a temperature contrast 
of a few degree K is found from equator to pole at the surface, the pole being warmer. In some observations a minimum 
at mid latitude with a warm equator and hotter polar regions is also found. The warm polar regions and cool equatorial region pattern
is also found in 3-D simulation of the solar convection zone with temperature variation slightly larger (i.e., of order 10 K; BT02, MBT06). 
At the surface a banded structure of the temperature field (warm-cool-hot) is also found in 3-D simulation of global scale convection. 
While very useful and instructive, most observations are confined to the solar surface and lack the information on the deep
thermal structure of the solar convection zone which is key to characterise the dynamics of the deep solar convection zone. 
One way to remedy that limitation is to rely on helioseismic inversions that allows us to probe deeper in the Sun and to use 3-D global simulations of the solar convection zone
to guide our physical understanding.

Indeed, helioseismic inversions can give us the rotation rate,
as well as the sound speed and density in the solar interior as a function
of radius and latitude. Inside the convection zone the chemical composition is
uniform and if we know the equation of state it is possible to determine
other thermodynamic quantities like the temperature and entropy from the
sound speed and density. Although, there may be some uncertainty in the
equation of state, the OPAL equation of state (Rogers et al.~1996;
Rogers \& Nayafonov 2002) is quite close to the equation of state of
solar material (e.g., Basu \& Antia 1995; Basu \& \jcd~1997). Thus, in this work we
use the OPAL equation of state to calculate the perturbations in entropy
and temperature and assess how well a strict thermal wind balance is established in the
solar convective envelope. To achieve this goal we make use of 2-D inversions of $\Omega, S, T$,
using the GONG and MDI data for the full solar cycle 23 and analyse our findings using 3-D
simulations obtained with the ASH (anelastic spherical harmonic) code (BT02; MBT06; Miesch et al.~2008) supported by 
theoretical considerations on the thermal wind balance and vorticity equations.

The rest of the paper is organised as follows: in Sect.~2 we describe
the data and technique used in this work while the results for the temperature and entropy inversions are
described in \S~3 along with those of 3-D simulations. In \S~4 we discuss at length the thermal wind balance and its
generalisation and interpret our seismic inversion with 3-D simulation of global scale convection. 
Finally, in \S~5 we put our results in perspective and conclude.

\begin{figure*}
\centerline{\resizebox{0.65\figwidth}{!}{\includegraphics{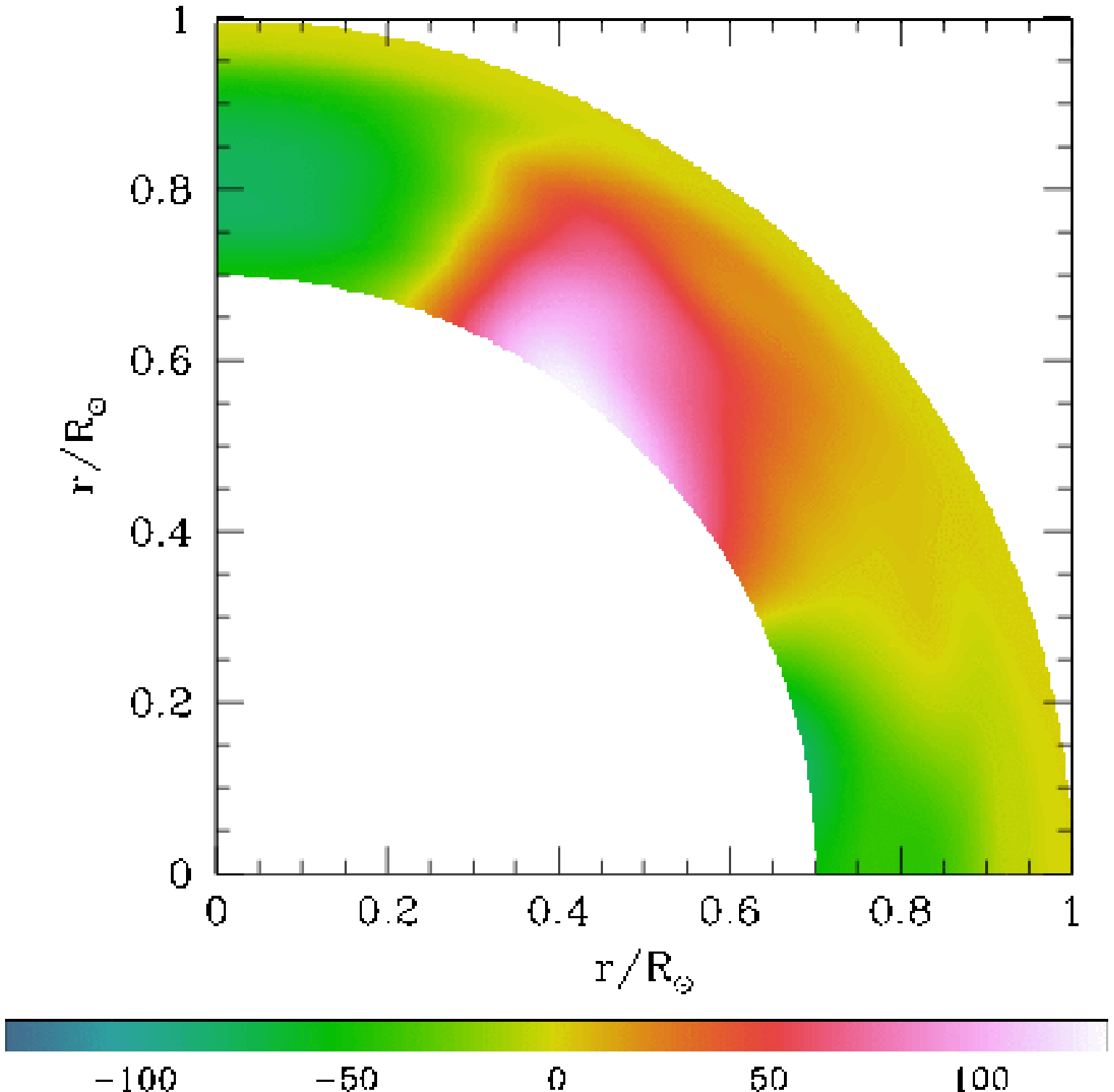} }\hfill
\resizebox{0.65\figwidth}{!}{\includegraphics{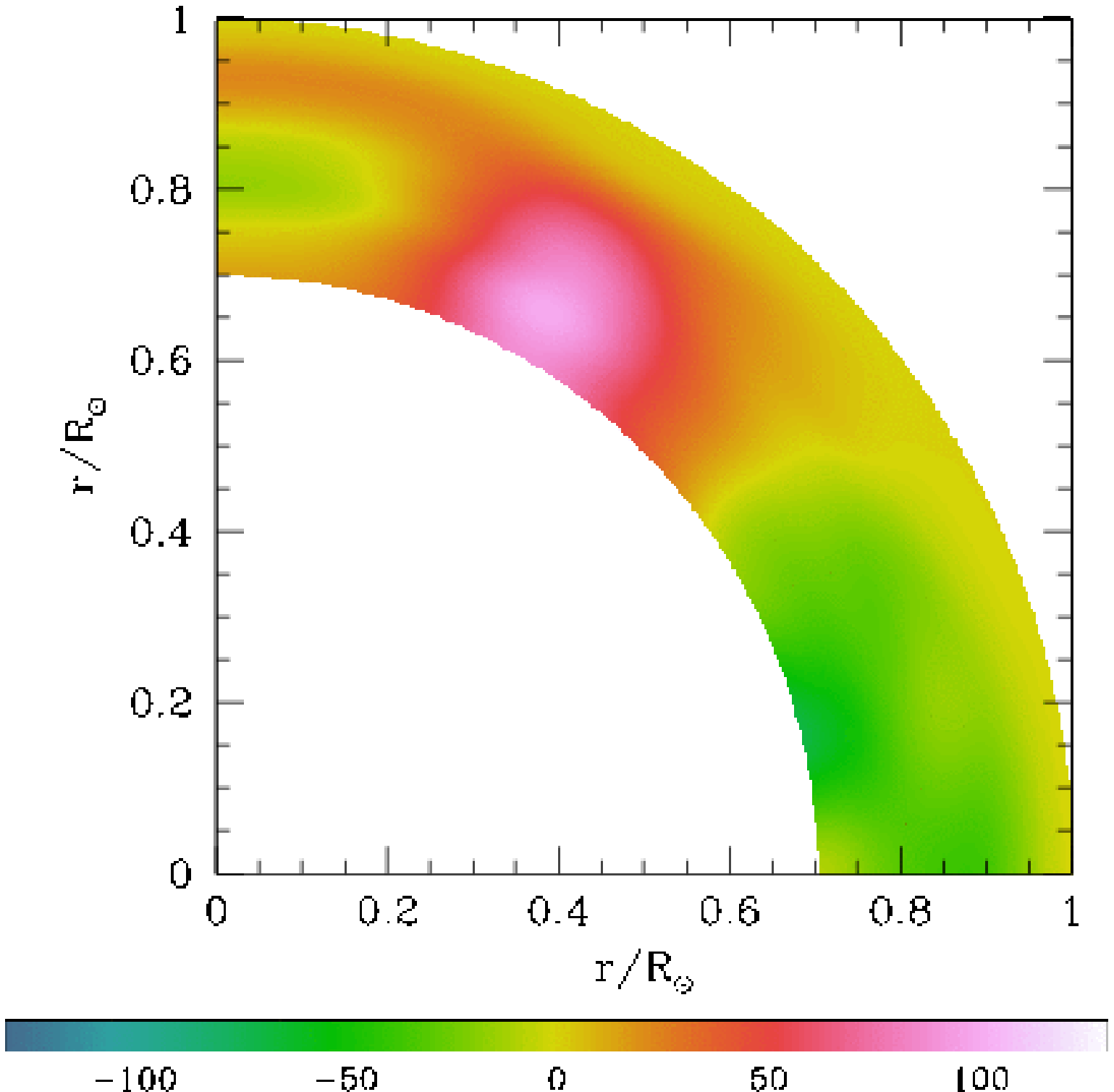} }\hfill
\resizebox{0.68\figwidth}{!}{\includegraphics{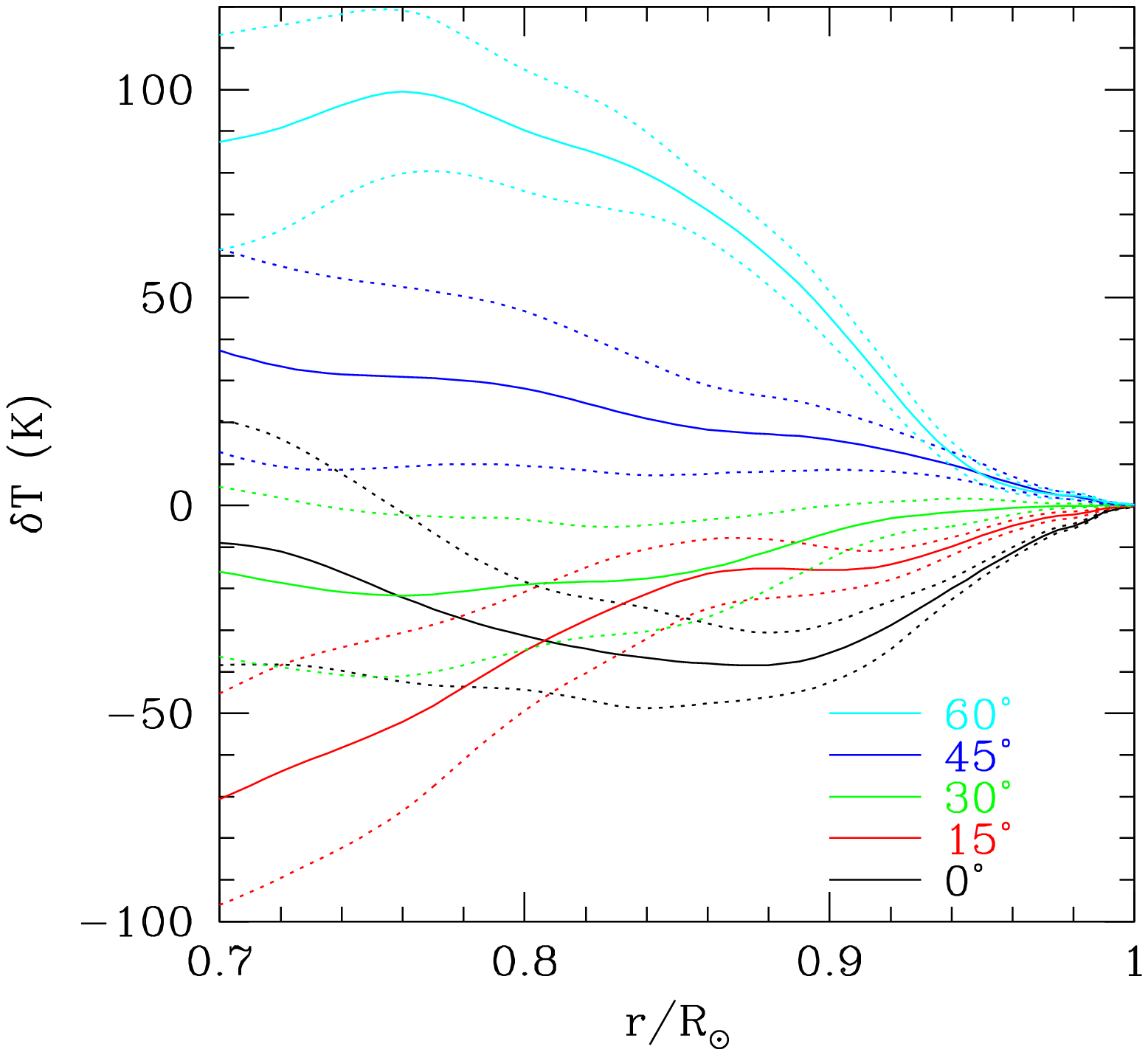} }}
\caption{The aspherical component of temperature fluctuation, $\delta T$
obtained from the temporally averaged GONG (left panel) and MDI (middle panel) data.
The right panel shows the cuts at constant latitude of $\delta T$ obtained
from MDI data along with $1\sigma$ error estimates shown by dotted lines.
All curves appear to merge
at $r=R_\odot$, because $\delta T$ is of order of 1 K in that region.}
\label{T_obs}
\end{figure*}

\begin{figure*}
\centerline{\resizebox{0.65\figwidth}{!}{\includegraphics{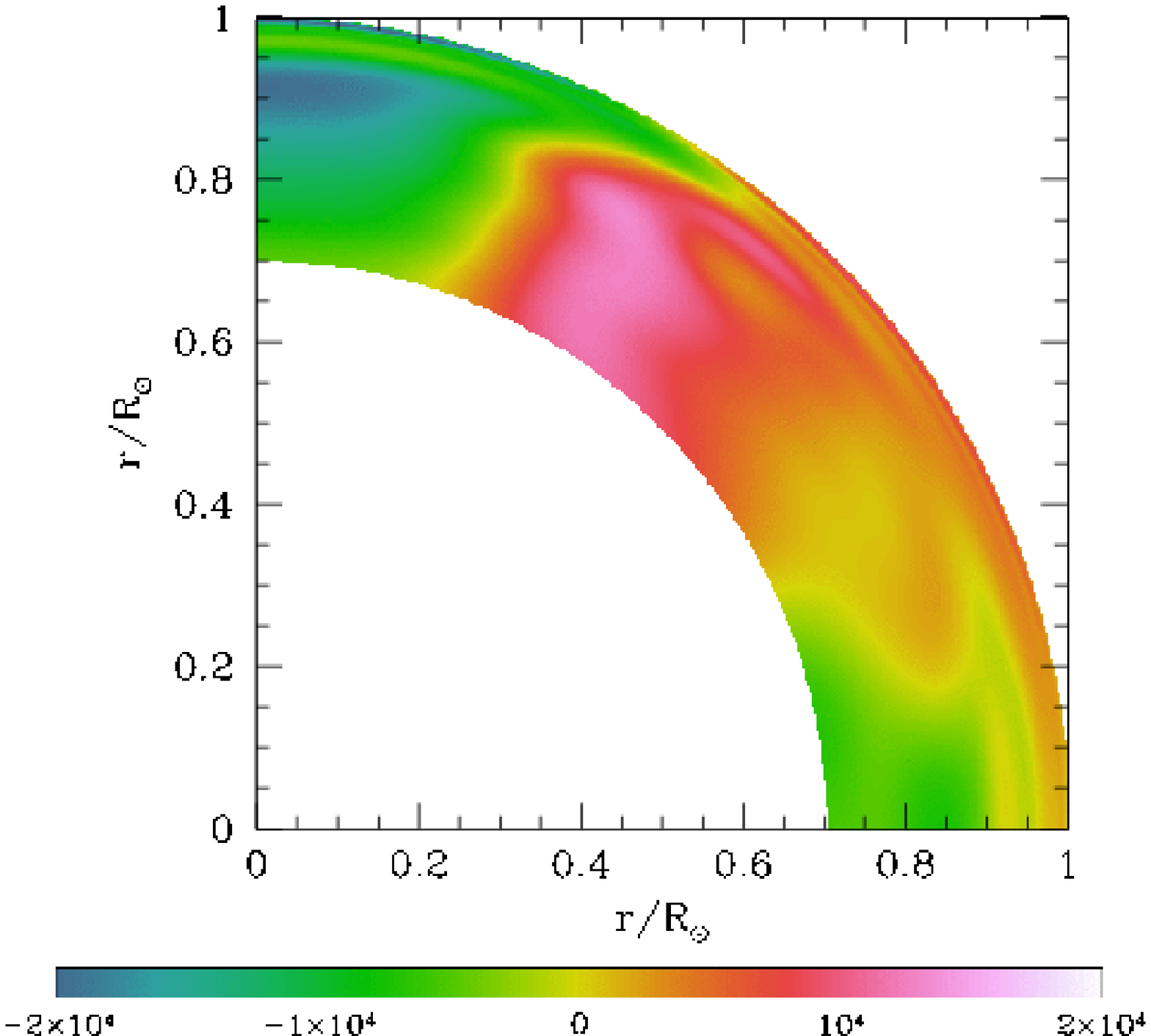} }\hfill
\resizebox{0.65\figwidth}{!}{\includegraphics{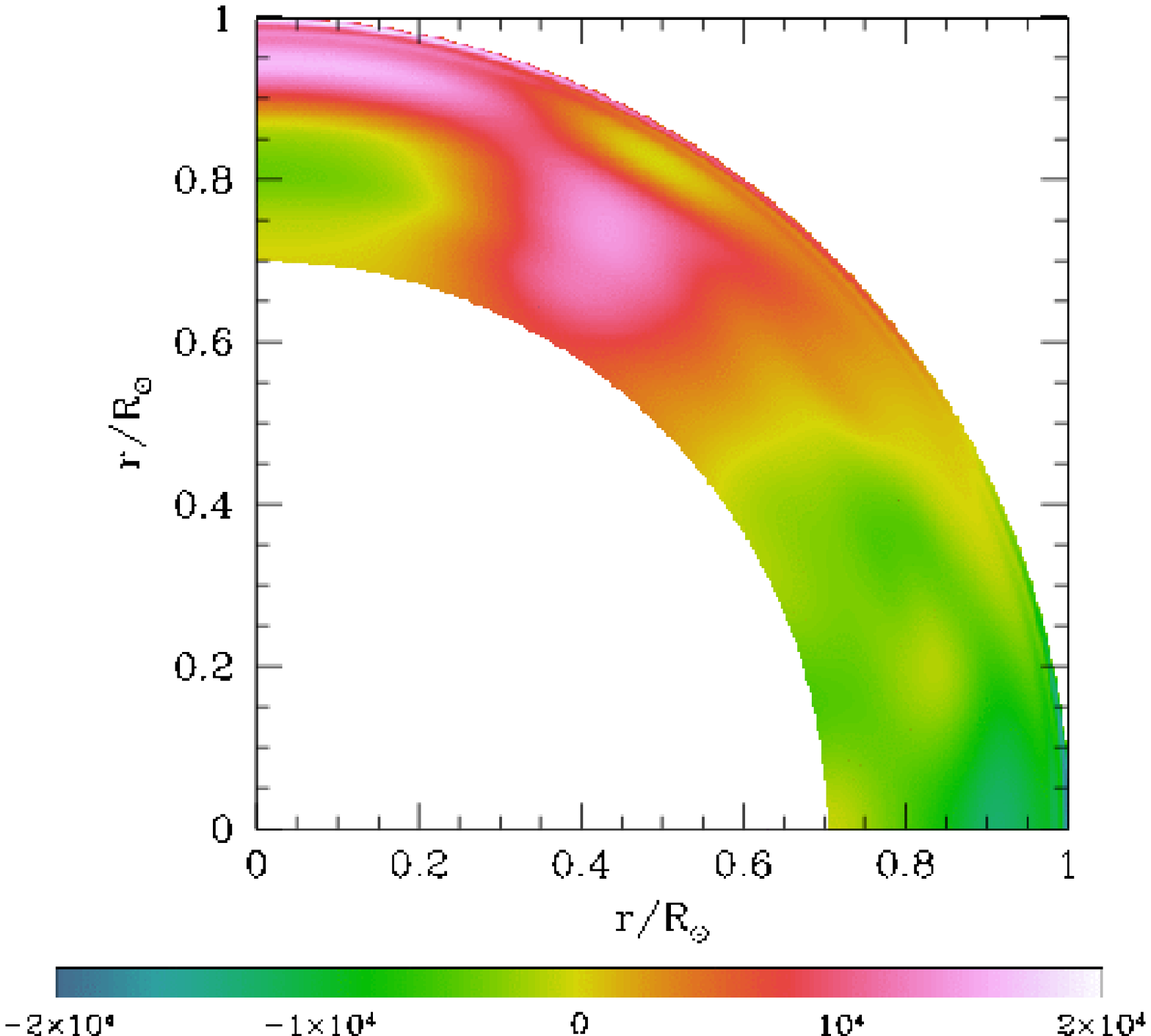} }\hfill
\resizebox{0.65\figwidth}{!}{\includegraphics{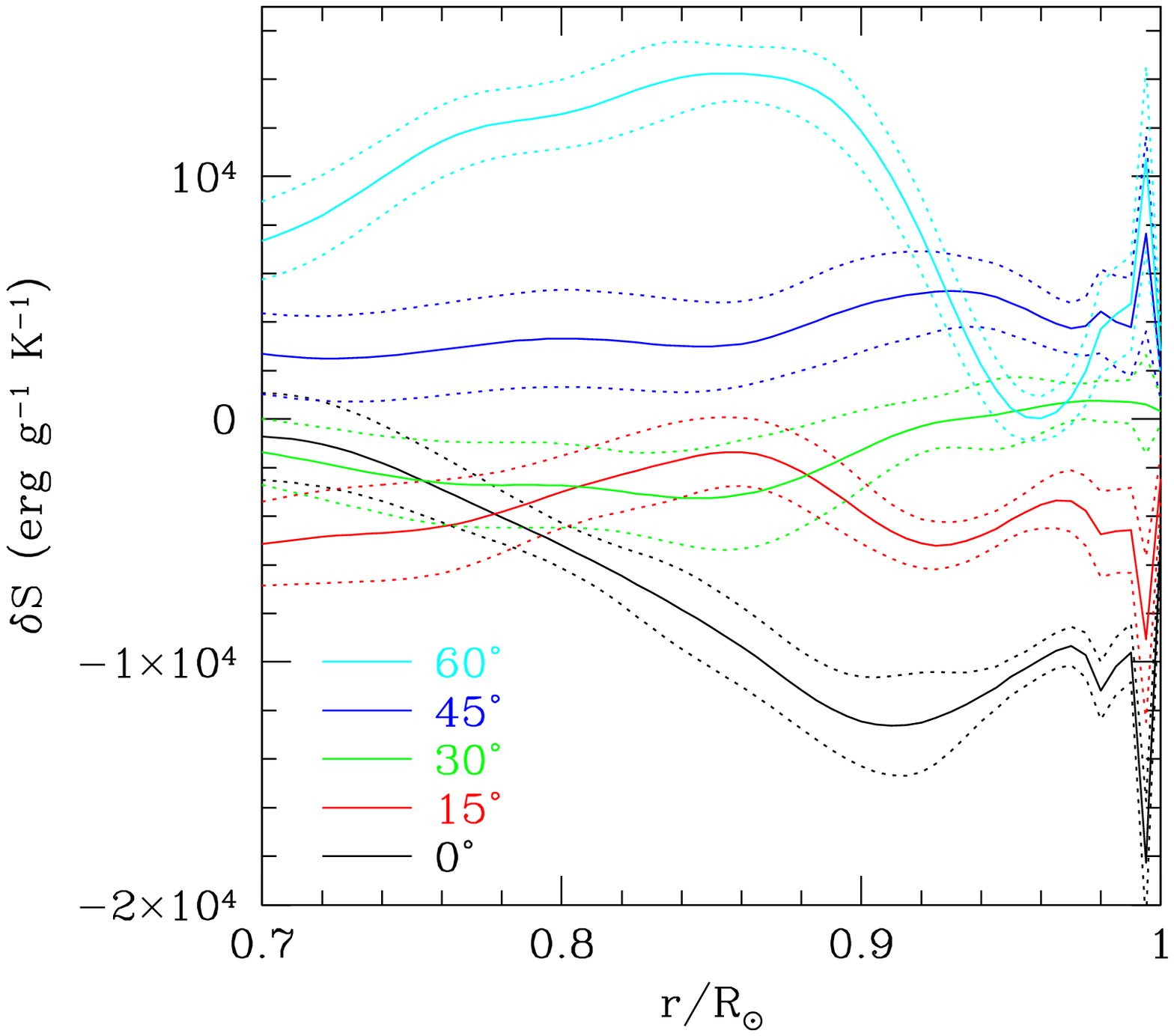} }}
\caption{The aspherical component of entropy fluctuation, $\delta S$
obtained from the temporally averaged GONG (left panel) and MDI (middle panel) data.
The right panel shows the cuts at constant latitude of $\delta S$ obtained
from MDI data along with $1\sigma$ error estimates shown by dotted lines.}
\label{S_obs}
\end{figure*}

\section{The helioseismic data and inversion technique}

We use data from GONG (Hill et al.~1996) and SOI/MDI (Schou 1999). Each data set
consists of mean frequencies of different $(n,l)$ multiplets,
and the corresponding splitting coefficients. We use 130 temporally overlapping
data sets from GONG, each covering a period of 108 days, starting
from 1995 May 7 and ending on 2008 May 9, with a spacing
of 36 days between consecutive data sets. The MDI data consist
of 61 non-overlapping data sets, each covering a period of 72 days,
starting from 1996 May 1 and ending on 2008 September 30. These data
cover the solar cycle 23. For most of the work we use the temporal average
over the available data to reduce the errors in inversion results.
For this purpose we repeat the inversion process for all data sets and
then take an average of all sets to get temporally averaged inversion
results.

We use a 2D Regularised Least Squares (RLS) inversion
technique in the manner adopted by Antia et al.~(1998) to infer the
angular velocity in the
solar interior from each of the available data sets.
Similarly, we use a 2D RLS inversion technique as described by
Antia et al.~(2001) to infer the sound speed and density in the solar
interior. In practice, we calculate the differences $\delta c^2/c^2$ and
$\delta\rho/\rho$ with respect to a reference solar model. We use the
solar model from Brun et al.~(2002) with tachocline mixing as the
reference model. In this work, we are only interested in the latitudinal
variation in solar structure inside the convection zone. Thus the 
fluctuation in sound speed can be converted to either temperature
or entropy using the relation:
\begin{eqnarray}
{\delta c^2\over c^2}&=&{\delta P\over P}+{\delta\Gamma_1\over\Gamma_1}
-{\delta\rho\over\rho}\\
\noalign{\medskip}
&=& \left(\left({\partial \ln P\over\partial \ln\rho}\right)_T+
{1\over\Gamma_1}\left({\partial\Gamma_1\over\partial\ln\rho}\right)_T -1\right){\delta\rho\over\rho}
\nonumber\\
&&\qquad+\left(\left({\partial \ln P\over\partial\ln T}\right)_\rho+
{1\over\Gamma_1}\left({\partial \Gamma_1\over\partial\ln T}\right)_\rho\right){\delta T\over T}\\
\noalign{\medskip}
&=& \left(\left({\partial \ln P\over\partial \ln\rho}\right)_S+
{1\over\Gamma_1}\left({\partial\Gamma_1\over\partial\ln\rho}\right)_S -1\right){\delta\rho\over\rho}
\nonumber\\
&&\qquad+\left(\left({\partial \ln P\over\partial S}\right)_\rho+
{1\over\Gamma_1}\left({\partial \Gamma_1\over\partial S}\right)_\rho\right)\delta S.
\end{eqnarray}
Here $S$ is the specific entropy, $T$ is temperature, $P$ is pressure and $\Gamma_1$ is
the adiabatic index. The required partial derivatives are calculated using
the OPAL equation of state. The derivatives of $\Gamma_1$ are small in
most of the convection zone, except for the ionisation zones of hydrogen
and helium, but for completeness we have included these derivatives in
all our calculations. Helioseismic inversions for rotation and asphericity
are only sensitive to the North-South symmetric components and hence the
inverted profiles always show this symmetry. Hence, in this work we show
the inversion results in only one hemisphere. Actual profiles may have
some asymmetry about the equator.

\section{Thermal perturbations in the solar convection zone}

Convection is a macroscopic transport of heat and energy. It is directly associated to correlation between the velocity field and temperature fluctuations (Brun \& Rempel 2008). 
Being able to infer the temperature and entropy perturbations in the solar convection zone is thus key to understanding its turbulent dynamics.

\subsection{The Inverted profiles}\label{ST_obs}

The aspherical part of temperature and entropy perturbations determined from
temporally averaged GONG and MDI data are shown in Figures \ref{T_obs} and \ref{S_obs}. The
maximum temperature fluctuation near the bottom of the convection zone is found to be about 100 K.
These fluctuations increase with depth initially, because of steep increase in
the temperature with depth which can induce an artificially large value for $\delta T$. 
The errors in $\delta T$ also increase with depth and the results may not be significant near the base of the convection zone.
If we consider the relative fluctuation $\delta T/T$, then the maximum would be much closer to the surface and the
value is of order of $10^{-4}$ or less. Similarly, if the entropy fluctuation
are divided by its typical value of order $C_p$, then it too would be of
the same order. Both these relative perturbations are of the same order as
$\delta c^2/c^2$. A detail look at Figures \ref{T_obs} and \ref{S_obs} reveal that the fluctuations
are negative (relatively cold with respect to the spherically symmetric mean) at low latitude and warm 
at mid latitudes. In the bulk of the solar convection zone there is very little radial variation except near the surface. 
In the GONG data a cool polar region is also apparent but  its significance is questionable given the relatively
poor resolution of inversion techniques at high latitude.
This feature is not clearly seen in the MDI data.
While this latitudinal variation imprints through the surface for
the entropy with little change in amplitude it is not the case for the temperature. At the surface the seismic inversion of 
the axisymmetric temperature fluctuations are very small in agreement with previous photospheric study (Rast et al.~2008).
It may be noted that near the surface, above the lower turning point of the
modes, the inversions may not be reliable. Around $r=0.95R_\odot$, where the
inversions should be reliable, the temperature variations are of the order of
10 K.

\begin{figure*}
\center
\includegraphics[width=0.55\linewidth,angle=90]{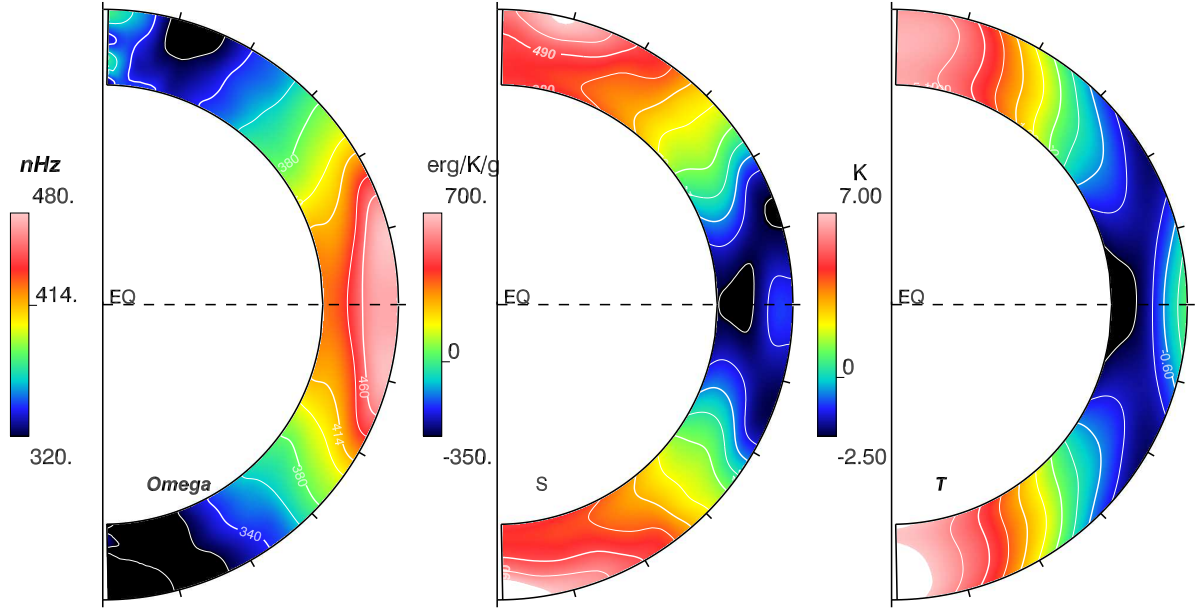}
\caption{Left panel: Angular velocity (in nHz) achieved in model AB3 of Miesch et al.~2006. Middle panel: Associated
entropy $S'$ (erg/g/K) and temperature $T'$ (K) fluctuations with respect to the spherically averaged background. All quantities have been
averaged over longitude and time (10 solar periods). Note the conical profile of the angular velocity at mid latitude and the 
latitudinal variation of the thermal variables possessing hot poles. Near the surface the temperature is banded with warm
equator, cool mid latitudes and hot poles. Contrary to the helioseismic inversion no symmetry with respect to the equator is assumed
and the quantities exhibit a small North-South asymmetry.}
\label{om_model}
\end{figure*}

\subsection{The profiles realised in 3-D models of large scale convection}

Recent efforts to develop high resolution global simulations of the solar convection zone in 
order to identify the physical processes at the origin of heat, energy and angular momentum transport have
been quite successful at reproducing the seismically inverted differential rotation profile (BT02; MBT06). 
We display in Figure \ref{om_model}, a typical solution of the solar convection zone and differential rotation obtained with the ASH code (case AB3 of MBT06). 
Shown using a meridional cut is the longitudinal and temporal average of the angular velocity $\Omega$ along with the temperature
and entropy fluctuations with respect to a spherically symmetric background. We first note that the differential rotation in the
model is solar-like, with a fast equator and slow pole, and iso-contours of $\Omega$ constant along radial lines
at mid latitude (i.e, the rotation profile is conical rather than cylindrical). Its amplitude is also of the right order of magnitude. By contrast it is important to note
that the temperature $T'$ and entropy $S'$ fluctuations\footnote[1]{for the sake of clarity we make the distinction between the 
seimic inversion of the temperature and entropy perturbations denoted with a $\delta$ symbol, and the one computed in the models denoted by a prime} are smaller by a factor of about 10 with respect to the seismic inversion, with temperature variations of
about 10 K from equator to pole up to $r=0.96 R_{\odot}$. A detailed analysis of the redistribution of heat and angular momentum in the 3-D models reveal that
the Reynolds stresses and the latitudinal enthalpy flux are key players in establishing the profile of angular velocity and the variation 
of temperature and entropy with latitude (Brun \& Rempel 2008). Reynolds stresses transport angular momentum from the polar region down to the equator being opposed
by meridional circulation and viscous effect. The heat is transported poleward by the turbulent enthalpy flux (e.g. $\bar{\rho} C_p \langle v'_{\theta} T' \rangle$, with $\langle \rangle$ denoting an azimuthal average, $\bar{\rho}$ the mean background density and $v'_{\theta}$ the fluctuating latitudinal component of the velocity field with respect to the axisymmetric mean, (see for more details Brun \& Palacios 2009)) yielding cool equator and hot poles in most of the domain. 
It is opposed by thermal diffusion that tries to make the entropy and temperature field homogeneous. A careful study of the profile of temperature and entropy fluctuations reveals that the
entropy is monotonic with respect to latitude while near the surface the temperature is banded (warm-cool-hot). Further the entropy profile is conical as is the angular velocity at mid latitude whereas the temperature 
profile is more cylindrical. In this stratified (anelastic) simulations the difference between the two thermal quantities is due to density (or pressure) fluctuations that cannot be neglected. 
This confirms that entropy is the key quantity to consider when studying the angular velocity profile of the Sun as clearly stated in the thermal wind equations detailed in \S 4.1.
Mean field 2-D models also find axisymmetric temperature variations of order few Kelvin 
at the surface and in the bulk of the convection zone (Kitchatinov \& R\"udiger 1995, K\"uker \& R\"udiger 2005). 
Current global 3-D numerical simulations of the solar convection zone do not model the very surface, but
stop at around 0.96 to 0.98 $R_{\odot}$, and as a consequence can not be used yet to model the near surface shear layer 
(see however the studies of Derosa et al.~(2002) using a modified ASH code or of Robinson \& Chan (2001), using a spherical wedge model).

\section{Quality of Thermal Wind Balance Achieved in the Sun and 3-D Models}

\subsection{Theoretical Considerations}\label{theory}

In rotating convection, both radial and latitudinal heat transport
occurs, establishing latitudinal gradients in
temperature and entropy within the convective zone as illustrated in Figure \ref{om_model}.
A direct consequence of the existence of such gradients is that the surfaces of pressure and density fluctuations will not coincide anymore,
thereby yielding baroclinic effects. We can turn to the vorticity equations
(\cite{Pedlosky87,Zahn92}) to analyse the role of the turbulence and baroclinic effects 
in setting the large scale flows shown in Figure \ref{om_model}. 
The thermal wind balance equation can be derived
from the vorticity equation as discussed in detail by BT02 and MBT06. The equation for the vorticity in the purely hydrodynamical
case can be derived under the anelastic approximation by taking the
curl of the momentum equation  (see also Fearn 1998 and Brun 2005 for its MHD generalisation and the 
notion of magnetic wind) :

\begin{eqnarray}\label{eq:vort}
\frac{\p \vort}{\p t}&= &(\vort_a\cdot\nab){\bf v} - ({\bf v}\cdot\nab)\vort_a - \vort_a(\nab\cdot{\bf v})\\\nonumber
 & + &\frac{1}{\rb^2}\nab\rb\times\nab P' - \curl\left(\frac{\rho g}{\rb}\uvr\right) - \curl(\rbi\nab\cdot\mbox{\boldmath $\cal D$}), 
\end{eqnarray}

with $\vort_a=\curl {\bf v} + 2\Om$ the absolute vorticity,
$\vort=\curl {\bf v}$ the vorticity in the rotating frame, and {\bf
  $\cal D$} the viscous tensor given by:
  
\begin{eqnarray}
{\cal D}_{ij}=-2\rb\nu[e_{ij}-\frac{1}{3}(\nab\cdot{\bf v})\delta_{ij}],
\end{eqnarray}
where $e_{ij}$ is the strain rate tensor, and $\nu$ is an effective kinematic viscosity. 

This vorticity equation helps in understanding the relative importance of
the different processes acting in the meridional planes. In the
stationary case ($\frac{\p \vort}{\p t} = 0$), and assuming an
azimuthal average (such that $\frac{\partial}{\partial \varphi}$ vanishes)
the azimuthal component of Eq.~(\ref{eq:vort}) reads:

\begin{eqnarray}\label{eq:TWfull}
2\Omega_0\frac{\p \langle v_{\phi}\rangle}{\p z}&=&\underbrace{-\langle (\vort\cdot\nab)v_{\phi} - \frac{\omega_{\phi}v_r}{r} - \frac{\omega_{\phi}v_{\theta}\cot\theta}{r}\rangle}_{\rm Stretching}\nonumber \\
& +& \underbrace{\langle({\bf v}\cdot\nab)\omega_{\phi} +\frac{v_{\phi}\omega_r}{r} + \frac{v_{\phi}\omega_{\theta}\cot\theta}{r}\rangle}_{\rm Advection} \nonumber\\ 
&-& \underbrace{\langle \omega_{\phi}v_r \rangle\frac{d\ln \rb}{d r}}_{\rm Compressibility} +\underbrace{\frac{1}{r}\left[\frac{\p}{\p r}(r \langle {\cal A}_{\theta}\rangle) - \frac{\p}{\p \theta}\langle {\cal A}_r \rangle \right]}_{\rm Viscous\; stresses} \\
&+& \underbrace{\frac{g}{r c_p}\frac{\p \langle S'\rangle}{\p \theta}}_{\rm Baroclinicity} + \underbrace{\frac{1}{r \rb c_p}\frac{d\bar{S}}{dr}\frac{\p \langle P'\rangle}{\p\theta}}_{\rm Non\; adiabatic\; stratification} \nonumber
\end{eqnarray}
where $\displaystyle \frac{\p}{\p z}=\cos\theta\frac{\p}{\p r}-\frac{\sin\theta}{r}\frac{\p}{\p \theta}$ and 
\begin{eqnarray}
%\begin{multline}
\langle{\cal A}_r\rangle&=& \rbi\langle\left[\frac{1}{r^2}\frac{\p(r^2{\cal D}_{rr})}{\p r}+\frac{1}{r\sin\theta}\frac{\p(\sin\theta{\cal D}_{\theta r})}{\p\theta} - \frac{ {\cal D}_{\theta \theta} + {\cal D}_{\phi \phi}}{r} \right]\rangle , \nonumber \\
\langle{\cal A}_{\theta}\rangle&=& \rbi\langle\left[\frac{1}{r^2}\frac{\p(r^2{\cal D}_{r\theta})}{\p r}+\frac{1}{r\sin\theta}\frac{\p(\sin\theta{\cal D}_{\theta \theta})}{\p\theta} \right] \\
&+& \rbi\left[ \frac{ {\cal D}_{\theta r} - cot \theta {\cal D}_{\phi \phi}}{r} \right]\rangle . \nonumber
%\end{multline}
\end{eqnarray}

\noindent In the above equation we have identified several terms:
\begin{itemize}
\renewcommand{\labelitemi}{$\bullet$}
\item $\rm Stretching$ describes the stretching $/$ tilting of the absolute vorticity due to velocity gradients;
\item $\rm Advection$ describes the advection of vorticity by the flow;
\item $\rm Compressibility$ describes the stretching of vorticity due to the flow compressibility;
\item ${\frac{g}{r c_p}\frac{\p \langle S'\rangle}{\p \theta}}$ is the baroclinic term, characteristic of non aligned density and pressure gradients;
\item $\frac{1}{r \rb c_p}\frac{d\bar{S}}{dr}\frac{\p \langle P'\rangle}{\p\theta}$ is part of the baroclinic term but arises from departure to adiabatic stratification;
\item $\rm Viscous$ accounts for the diffusion of vorticity due to viscous effects. 
\end{itemize}
We wish to stress that for the nonlinear stretching and advection terms (equivalent to Reynolds stresses in Navier-Stokes equation) 
their azimuthal average still yields partial derivatives in $\phi$, since quadratic terms such as $\langle \frac{\omega_{\phi}}{r \sin \theta}\frac{\p v_{\phi}}{\p\phi}\rangle$ are non zero.\\

Under the assumption that the convection zone is adiabatic, the Rossby number $R_o = \omega/2\Omega_0$ is small, and that compressibility, Reynolds and viscous stresses can be neglected, 
equation (\ref{eq:TWfull}) simplifies to give: 

\begin{equation}\label{eq:TW}
\centering
\frac{\p \langle v_{\phi}\rangle}{\p z}=\frac{g}{2 \Omega_0 r c_p}\frac{\p \langle S'\rangle}{\p \theta}
\end{equation}
This is the thermal wind equation. It simply states that baroclinic effect can break Taylor-Proudman constraint of cylindrical differential rotation since otherwise $\p v_{\phi}/{\p z}=0$
for barotropic flows (Zahn 1992). 
This is due to the fact that the baroclinic terms drive meridional flows
that under the influence of Coriolis force yield longitudinal flows that lead to a non cylindrical state of rotation.  We now turn to our numerical
simulation to evaluate the role played by all the terms of the vorticity equation identified above and to discuss the quality of the thermal wind balance achieved.

\subsection{Results from 3-D Models}

\begin{figure}
\includegraphics[width=1.45\linewidth]{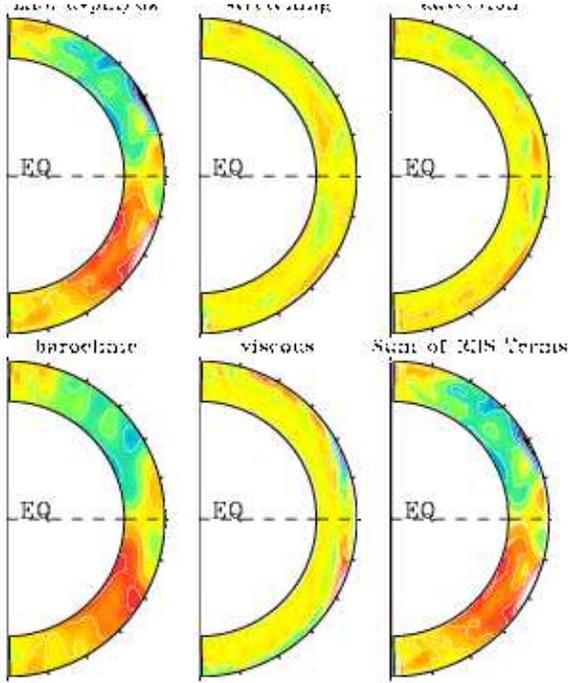}
\caption{Meridional cut of the terms discussed in equation (6) averaged over azimuth and 10 solar periods.
Shown are in turn: $\p \langle v_{\phi}\rangle/\p z$, the stretching and advection of vorticity, the baroclinic effects,  the viscous stresses and the sum of the RHS terms (we have divided all the RHS terms
by $1/2\Omega_0$). Each panel share the same colour table with red denoting positive value. The min/max values used to scale the plots are $[-10^{-6}, 10^{-6}]$ expressed in $s^{-1}$.}
\label{TW_theo}
\end{figure}

Figure~\ref{TW_theo} displays for case AB3 the left-hand side of
Eq.~(\ref{eq:TWfull}), along with the dominant terms of the right-hand side and their sum.
We clearly see that the sum of the dominant RHS term is
in very close agreement with the LHS.  We have chosen to form the temporal average over 10 solar periods because it corresponds to about
10 convective overturning times and lead to a very close balance between the LHS and the RHS of equation (\ref{eq:TWfull}). Shorter averages do not
lead to such a good balance whereas longer averages do not change the quality of the balance obtained significantly nor the patterns of the various terms. 
Our more detailed decomposition of the vorticity equation is allowing us to
identify which term is contributing and where. First the baroclinic term is found to be dominant in most of the bulk of
the convection zone as was found by BT02 and MBT06. Advection terms are found to contribute both in the bulk and near the surface. They
do not possess contrary to the baroclinic term a systematic dominant contribution in each hemisphere. Their contribution leads to change in key places, yielding
a more structured profiles of the RHS than the baroclinic term would have yielded if considered alone. Since the Rossby number realised in the simulation is less than one, we expect the
advection and stretching term to be small on average in the simulation and indeed their maximum amplitude is not as large as the baroclinic term. However as stressed above 
this is not the case at all scales nor at all locations  and they do contribute in key places,  leading to the very good balance shown in Figure~\ref{TW_theo} between the LHS and RHS of equation~(\ref{eq:TWfull}).
Finally, in our models a viscous shear layer is dominating the
balance at the surface where the isocontours of $\Omega$ possess the strongest latitudinal shear. Durney (1989) and Kitchatinov \& Ruediger (1999) have also stressed
that a strict thermal wind balance cannot be realised everywhere in the convection zone and that viscous stresses may play a role near the boundaries as observed in Figure~\ref{TW_theo}.
We can thus conclude that equation (\ref{eq:TW})
 is only partly satisfied in our 3-D hydrodynamical simulations of the solar convective envelope. Clearly baroclinic effects play a central role but these are far from being 
 dominant everywhere and considering only equation  (\ref{eq:TW}) instead of the full balance expressed in equation (\ref{eq:TWfull}), would be misleading.  We now turn to seismic inversion to see if
 the thermal wind balance is strictly realised in the Sun or if other contribution must be invoked to explain the
 solar peculiar rotation profile.

\subsection{Inverted Solar Thermal Wind Balance}

\begin{figure*}
\centerline{\resizebox{\figwidth}{!}{\includegraphics{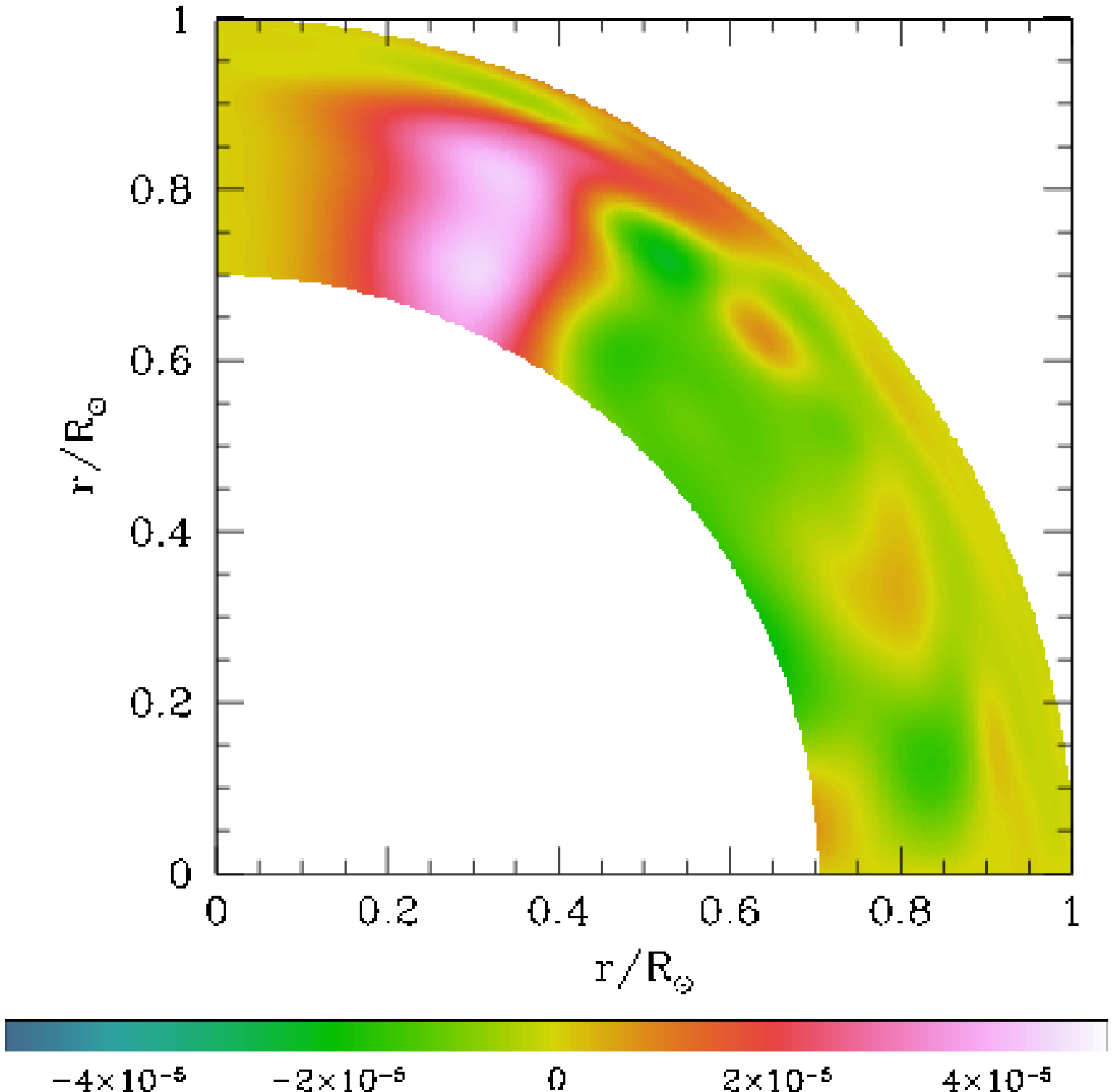} }\hfill
\resizebox{\figwidth}{!}{\includegraphics{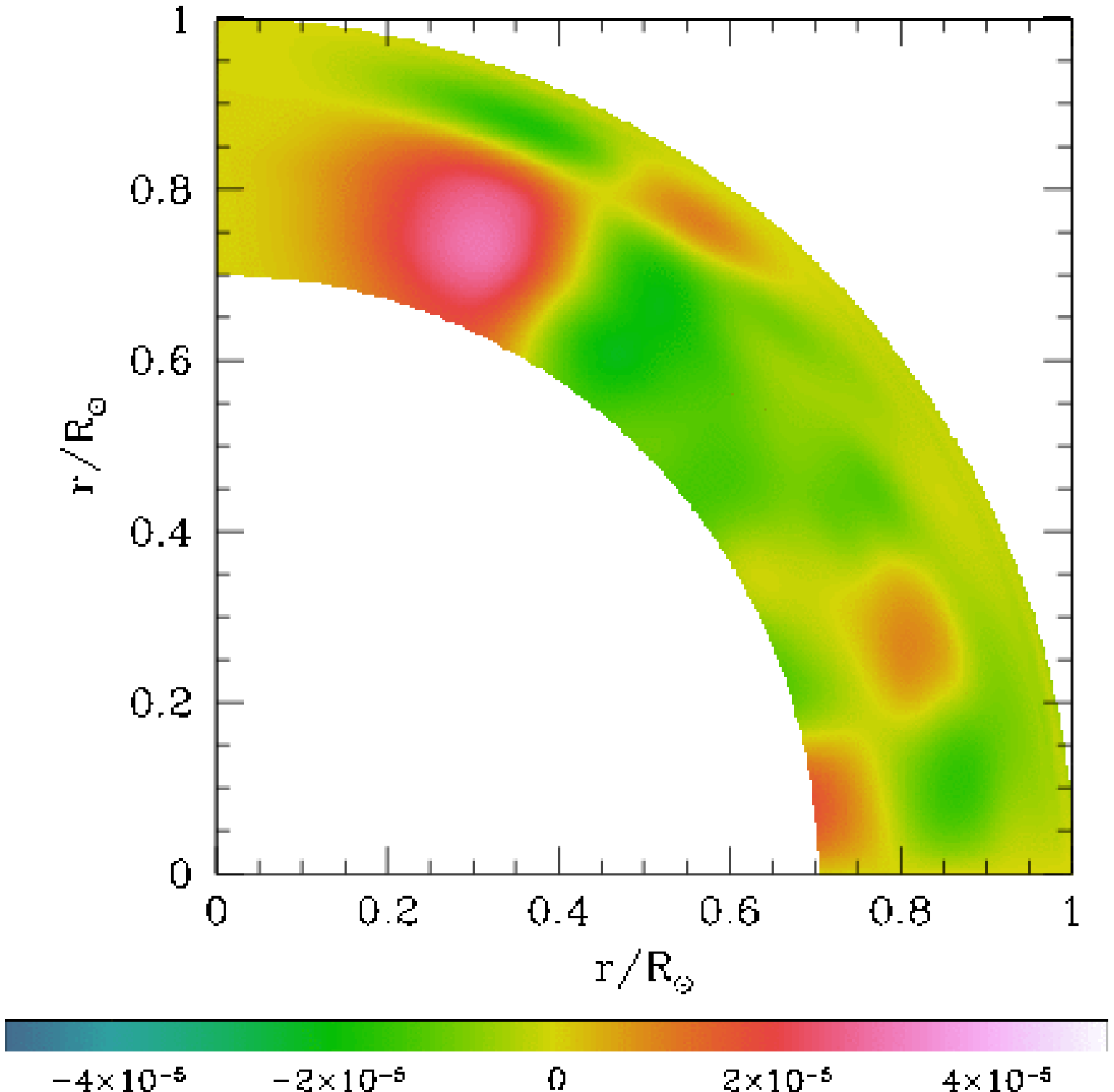} }}
\caption{The aspherical component of the latitudinal derivative of
the entropy fluctuation, $(g/2\Omega_0 rC_p)(\partial \delta S/\partial\theta)$
obtained from the temporally averaged GONG (left panel) and MDI (right panel) data. The values are in $s^{-1}$.}
\label{TW_obs}
\end{figure*}

\begin{figure*}
\centerline{\resizebox{\figwidth}{!}{\includegraphics{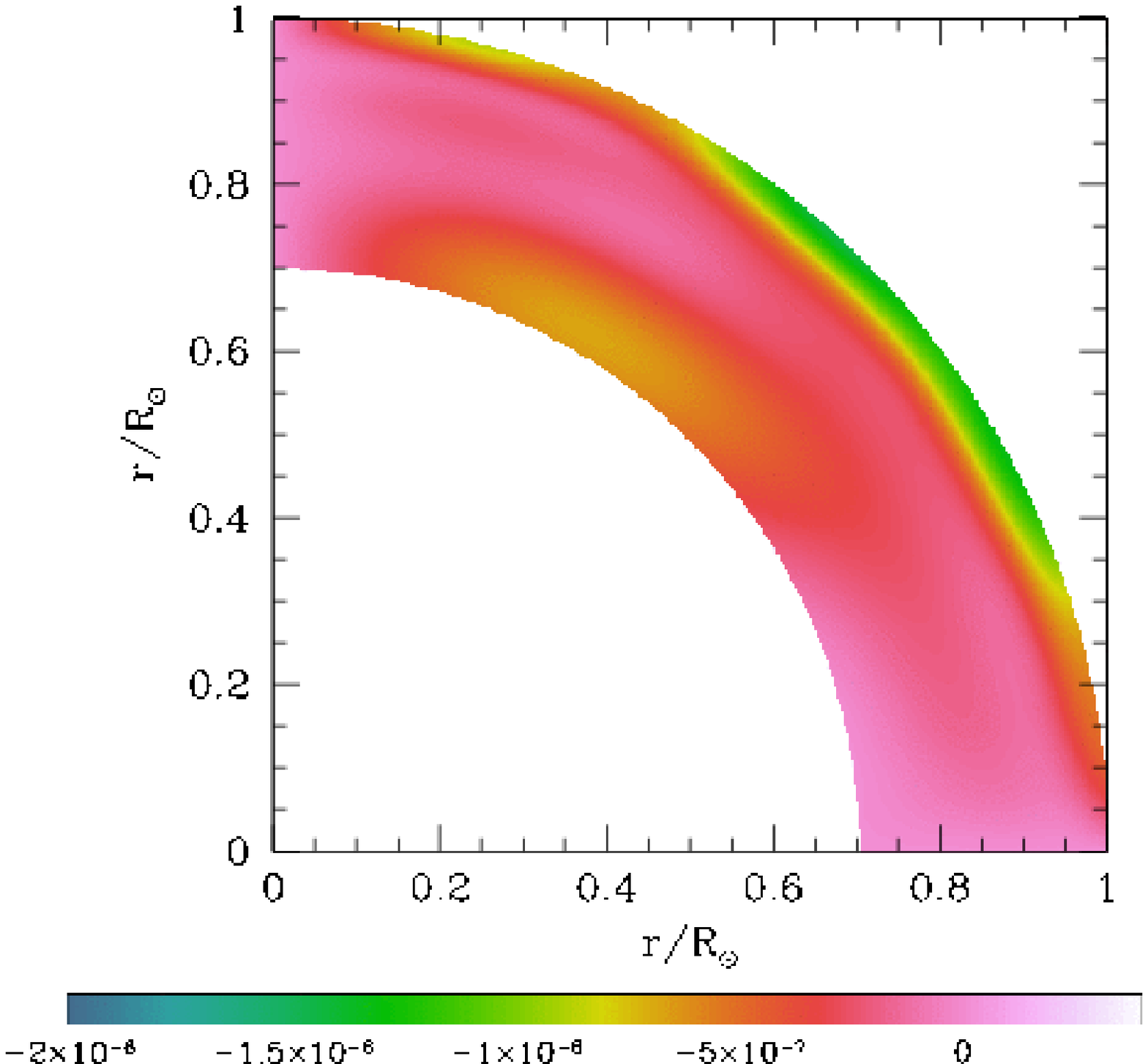} }\hfill
\resizebox{\figwidth}{!}{\includegraphics{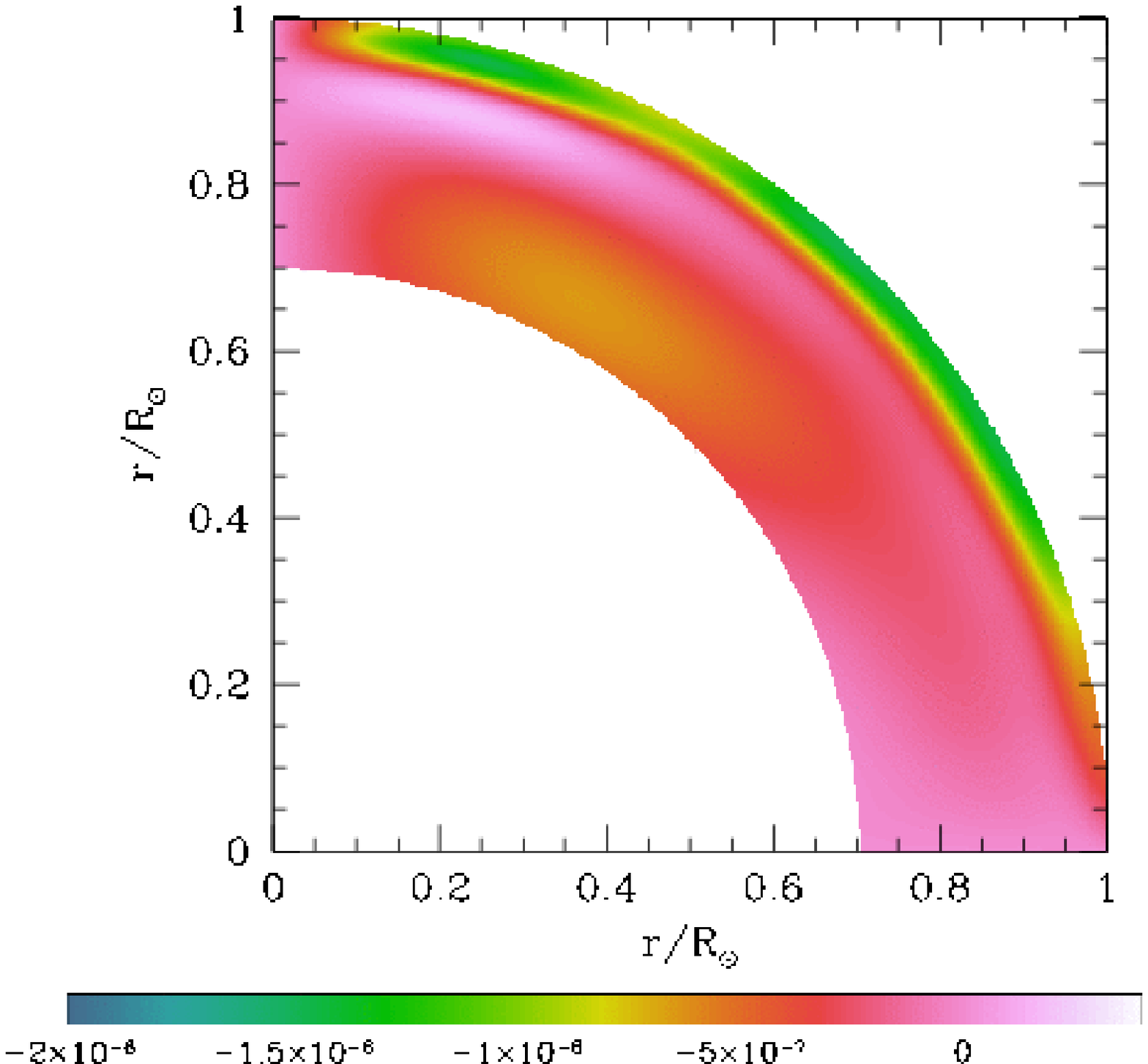} }}
\caption{The derivative of rotation velocity,
$\sin\theta(r\cos\theta \partial\Omega/\partial r -\sin\theta
\partial\Omega/\partial\theta)$,
obtained from the temporally averaged GONG (left panel) and MDI (right panel) data. The values are in $s^{-1}$.}
\label{dOm_obs}
\end{figure*}

The entropy perturbations obtained in section \ref{ST_obs} can be differentiated to calculate the RHS of equation (\ref{eq:TW})
\begin{equation}
\centering
{g\over 2\Omega_0 r C_p}{\partial \delta S\over \partial\theta}.
\end{equation}
The result is shown in Figure \ref{TW_obs}.  We clearly see that the baroclinic term is non monotonic with respect to
latitude, with large positive values near the poles and in a small region at the equator whereas it is negative in mid latitudes.
At the surface a surface thermal boundary layer is visible that yields strong radial gradients at high latitudes.

As we have done with the 3-D model, the baroclinic term should be compared with
\begin{equation}
\centering
\sin\theta\left(r\cos\theta {\partial\Omega\over\partial r} -\sin\theta
{\partial\Omega\over\partial\theta}\right),
\end{equation}
which is shown in Figure \ref{dOm_obs}. This quantity has much less structure in the bulk of the convection zone.
Except for a slightly negative structure at mid depth and latitude, most of the action occurs in the surface shear layer where strong negative
values are found due to strong radial gradient of rotation rate in the
near surface shear layer. This near surface layer is not present in the simulations and hence
can not be compared with the results of 3-D simulations.
It is clear that contrary to what we have seen with the 3-D model in the previous section, the two quantities do not agree with each other even slightly. 
In fact, these two terms differ by more than an order of magnitude. While the term involving $\Omega$ in equations 8 or 10 is of the same order in both
the 3-D simulations and the seismic inversion ($\sim 10^{-6}\, s^{-1}$), this is not the case for the baroclinic terms due to the very large entropy 
and temperature variations in the inverted profiles. Although the $\Omega$ profile in simulations qualitatively reproduces the
features seen in solar profile, the detailed latitudinal variations in the two do not match precisely.

\section{Discussion of Results}

%{\bf please extend/comment}\\
What can be the source of the disagreement between the inverted baroclinic contribution and the $z$ derivative of the angular velocity (i.e equations 8, or 9 and 10)?
The first and easiest solution is that the inversion of the thermal quantities lack the necessary accuracy and given the increase by two orders of magnitude of the background temperature
and density with depth, we end up with variations that are too large. The source of discrepancy will then be due to an overestimation of $\delta T$ and $\delta S$. It is not easy to 
decide if these inverted thermal fluctuations are too large or if the simulations (both 2-D and 3-D) underestimates
the fluctuations realised in the Sun, because for instance of their limited Reynolds number. We must thus also consider the possibility that these large thermal perturbations are genuine. If this is
indeed the case we need to see how we could resolve the discrepancy between the seismically inverted LHS and RHS of equation 8.
As stated in section \ref{theory}, to obtain a strict thermal wind balance as expressed in equation 8, one need to make a certain number of assumptions:
adiabaticity, weak Rossby number, negligible compressibility, viscous and Reynolds stresses, stationarity. Further by considering only the hydrodynamic contributions we have omitted 
those associated with Maxwell stresses that are certainly present in the magnetic Sun. Let's however assume that the Maxwell stresses are not the source of the large observed 
discrepancy. We are confident that this is the case because we have formed temporal averages over a maximum and minimum period of activity and the differences
between the two periods are about 10 times smaller that what it would required if all the sources of discrepancy were coming from the Maxwell stresses alone. We nevertheless intend to make a more systematic study of the departure
of the strict thermal wind balance linked to magnetic effects (i.e. via the so called magnetic wind) by analysing the solar cycle 23 in details and by comparing with dynamo simulations of the solar convection (Brun et al.~2004).
We must thus question the validity of the other hypothesis made in deriving equation 8. Clearly it is justified given the very low microscopic value of the solar kinematic viscosity to consider that the viscous terms
do not contribute much. This is clearly not the case in the 3-D models where near the surface they are major contributors to the overall balance (see Figure \ref{TW_theo},
middle panel of the bottom row), but this is due to our large effective viscosity. Assuming adiabaticity is certainly reasonable in most of the convection zone
but clearly not near the surface. Since we are mostly interested in understanding the bulk dynamics of the solar convection zone, this term is indeed very small.
The choice of low Rossby number that allows us to neglect $\omega$ over $2\Omega_0$ 
the planetary vorticity is certainly not justified at all scales of the turbulent velocity spectra, in particular for those scales much smaller than the Rossby radius of deformation 
(Pedlosky 1987). In the Sun the large range of convection scales certainly undergo different dynamics depending on how sensitive they are to the Coriolis force.
The subtle angular momentum and heat redistribution realised in the  Sun is in part captured in our 3-D models. We can thus analyse if the Reynolds stresses associated with the turbulent
motion indeed play a central role. As discussed in detail in Brun \& Toomre (2002) and in \S 4 we know that it is indeed the case in our numerical simulations (see Figure \ref{TW_theo}, middle and right panel of the top row) 
even though our simulation do not possess a Reynolds number and a degree of turbulence as high as that in the Sun. We can thus expect, given the very large Reynolds number of the solar convection zone, 
that Reynolds stresses must play a central role in the Sun in shaping the differential rotation profile and that they somehow in part compensate the baroclinic 
contribution to yield the observed profile of angular velocity. This is a very interesting results, since
 it indicates that the differential rotation is of dynamical origin for its equatorial acceleration 
 (as revealed by studying angular momentum transport in our simulations as in BT02 or Miesch et al.~2008) but also most certainly for its shape  with Reynolds stresses helping or opposing in some
 regions the baroclinic effects to break Taylor-Proudman constraint. Of course this conclusion only holds if the inverted large thermal fluctuations are real.

\begin{acknowledgements}
We thank J.P. Zahn for useful comments on a draft version of this paper.
We acknowledge funding by the Indian-French scientific network (IFAN). One
of us (A.S.B.) is grateful to the Tata Institute of Fundamental Research, Mumbai and the Indian Institute of Astrophysics, Bangalore 
and its director Prof. S. Hasan for their hospitality during his visit in November 2008.
This work  utilised data obtained by the Global Oscillation
Network Group (GONG) project, managed by the National Solar Observatory,
which is
operated by AURA, Inc. under a cooperative agreement with the
National Science Foundation. The data were acquired by instruments
operated by the Big Bear Solar Observatory, High Altitude Observatory,
Learmonth Solar Observatory, Udaipur Solar Observatory, Instituto de
Astrofisico de Canarias, and Cerro Tololo Inter-American Observatory.
This work also utilises data from the Solar Oscillations
Investigation/ Michelson Doppler Imager (SOI/MDI) on the Solar
and Heliospheric Observatory (SOHO).  SOHO is a project of
international cooperation between ESA and NASA. A.S.B  acknowledges funding
by the European Research Council through ERC grant STARS2 207430 (www.stars2.eu).

\end{acknowledgements}

\end{document}